\documentclass[12pt]{iopart}

\newcommand{\be}{\begin{equation}}
\newcommand{\ee}{\end{equation}}

\newcommand{\ber}{\begin{eqnarray}}
\newcommand{\eer}{\end{eqnarray}}

\usepackage{iopams}  
\usepackage{graphicx}
\usepackage{epsfig}

\begin{document}

\title{Is the isoscalar monopole resonance of the $\alpha$-particle a collective mode?}

\author{G. Orlandini$^{1,2}$, S. Bacca$^{3,4}$, N. Barnea$^5$, W. Leidemann$^{1,2}$}

\address{$^1$ Dipartimento di Fisica, Universit\`a di Trento, Via Sommarive, 14 I-38123 Trento, Italy}

\address{$^2$ Istituto Nazionale di Fisica Nucleare, TIFPA, Via Sommarive, 14 I-38123 Trento, Italy}

\address{$^3$ TRIUMF, 4004 Wesbrook Mall, Vancouver, British Columbia V6J 2A3, Canada}

\address{$^4$ Department of Physics and Astronomy, University of Manitoba, Winnipeg, Manitoba R3T 2N2, Canada}

\address{$^5$ Racah Institute of Physics, Hebrew University, 91904, Jerusalem, Israel}

\begin{abstract}
In this contribution we review  and clarify the arguments which might allow the interpretation of 
the isoscalar monopole resonance of $^4$He as a collective breathing  mode.
\end{abstract}

%
%
\section{Introduction}
\label{intro}
The observable of interest in this contribution is the spectrum of  $^4$He  when it responds to
isoscalar monopole excitations. 
Such a spectrum  can be measured by perturbative isoscalar probes, like for example a beam of $\alpha$-particles. 

At the end of the 60's inclusive electron scattering experiments~\cite{FrR65,Wa70}  put in evidence the existence
of a narrow peak in the spectrum of $^4$He. This was ascribed just to an isoscalar monopole resonant excitation 
(0$^+$ resonance) and the transition form factor 
to this resonant state was measured for different momentum transferred $q$ up to 2 fm$^{-1}$. 
No measurement of the whole isoscalar monopole spectrum, however, is available for $^4$He.

In the case of larger systems, such spectra have been the object 
of considerable activity, both from the experimental and theoretical points of view. The isoscalar monopole spectra of these nuclei
exhibit visible ''bumps``, called  isoscalar giant monopole resonance (GMR).
The interest in these GMR's lies in the attempt to get an extrapolated value for the nuclear matter compressibility, 
a quantity of great astrophysical interest.

One has to notice that GMR's are visible in experiments at low momentum transfer   
and that the low energy part of the spectrum gives the main contribution to the 
compressibility. 
Of course, since for low $q$ the wavelength that probes the target is large
the spectrum will show mainly features that involve all constituents and therefore can reveal collective behaviors.
Therefore the GMR's have been interpreted as signatures  of ''breathing modes``.

In an attempt to bridge few- and many-body physics we think it is interesting to ask the question whether 
also the measured $0^+$ excited state in $^4$He might be interpreted as a ''collective`` state. A similar question arose in
Ref.~\cite{BaM02}, when the calculated dipole photonuclear cross section of $^6$He surprisingly exhibited two prominent peaks.  
That result suggested  a possible collective interpretation: the higher energy peak could have been ascribed to the displacement 
of the proton sphere against the neutron one (Gamow-Teller mode), that at lower energy to 
the displacement of the halo neutrons against a possible $\alpha$-core.

The $\alpha$-particle is a very compact system, with a similar binding energy per particle
as heavier nuclei, and it is often considered the nucleus where few- and many-body methods can be benchmarked.
Therefore in order to answer the question of the title, we apply to $^4$He the same criteria as those, 
present in the many-body literature,   
used to judge the collectivity  of ''bumps'' in the spectra.
They consist in studying both the moments of the spectrum (sum rules) and the transition densities.

For an isoscalar monopole spectrum $S_M(q,\omega)$ the moments and relative sum rules are  
\begin{equation}
 m_k(q)=\int \,d\omega \,\omega^k S_M(q,\omega)=\langle 0| M(q) \,H^k \,M(q) |0\rangle\,,
\end{equation}
where 
\begin{equation}
M(q)=\frac{1}{2}\left( \sum_i j_0(q r_i) - \langle 0|\sum_i j_0(q r_i)|0\rangle\right)\,.
\end{equation}
For $k\geq0$ these quantities can in principle  be calculated avoiding the knowledge of scattering states for energies in the continuum.

Transition densities to specific states $|n\rangle$ are defined as
\begin{equation}
\rho_{tr}(\vec r ) =\langle n| \hat \rho (\vec r) |0\rangle\,.
\end{equation}
\begin{figure}[htb]
\centering
\includegraphics[width=7cm,clip]{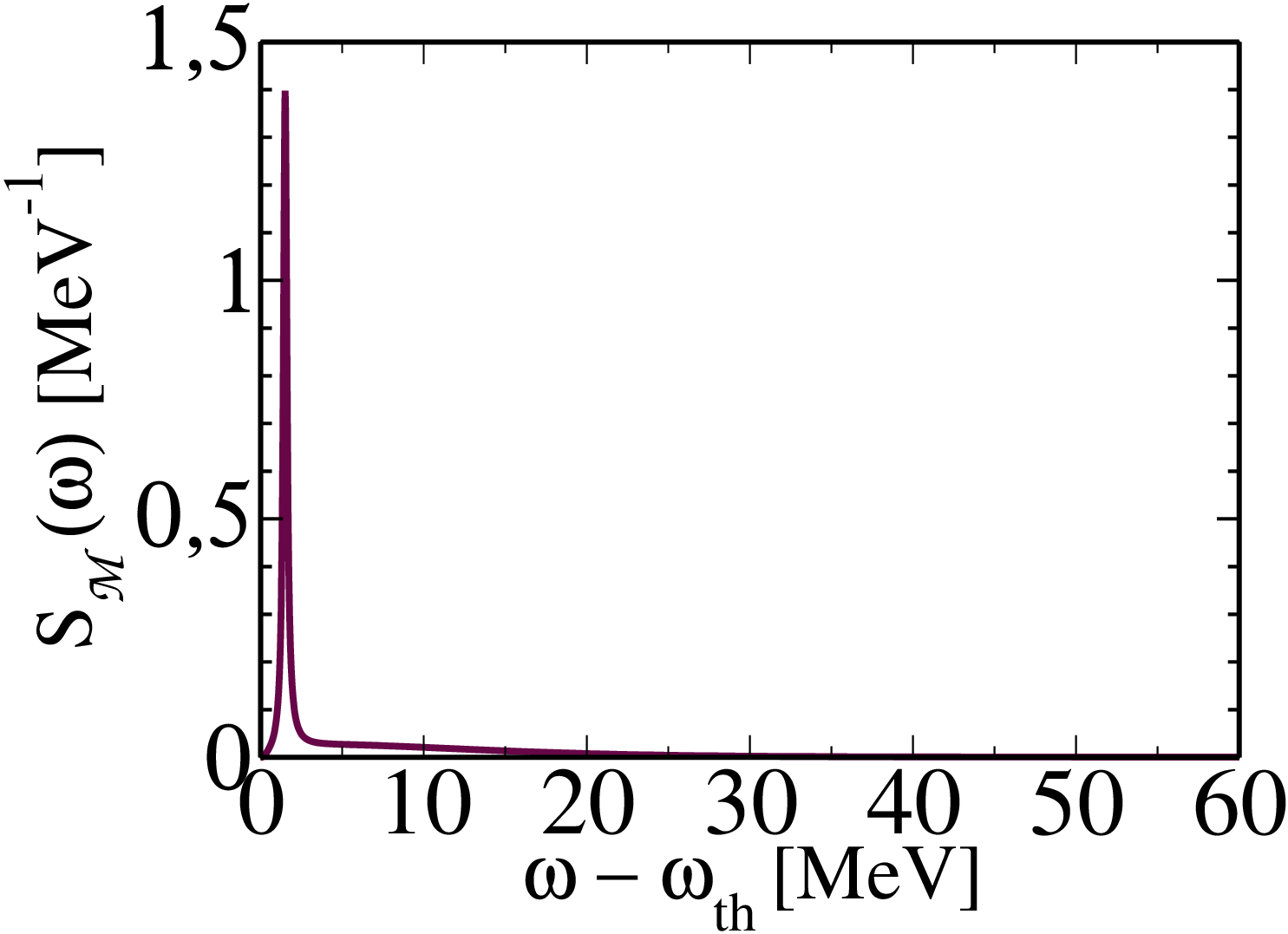}
\caption{The inelastic isoscalar monopole strength of $^4$He 
as a function of the excitation energy relative to threshold, for momentum transfer $q=0.25$ fm$^{-1}$.}  
\label{fig-1}       
\end{figure}
If the $0^+$ state in $^4$He were an excitation of extremely collective character the strength would have 
a $\delta$-function character. Therefore the strength of the resonance would 
exhaust all $m_k$. If this excitation were collective only to some degree, the ratio between the strength 
of the resonance and $m_0$ would give a measure of the {\it degree of collectivity}. Other sum rules for higher 
$k$ would not be fit for the purpose, since they emphasize too much the spreading of the  background strength to higher energy.
Moreover, if the $0^+$ state in $^4$He were an excitation of extremely collective character, described as a breathing mode,
the transition density would be zero at a value of $r$ equal to the radius of the system.

\begin{figure}
\centering
\includegraphics[width=7cm,clip]{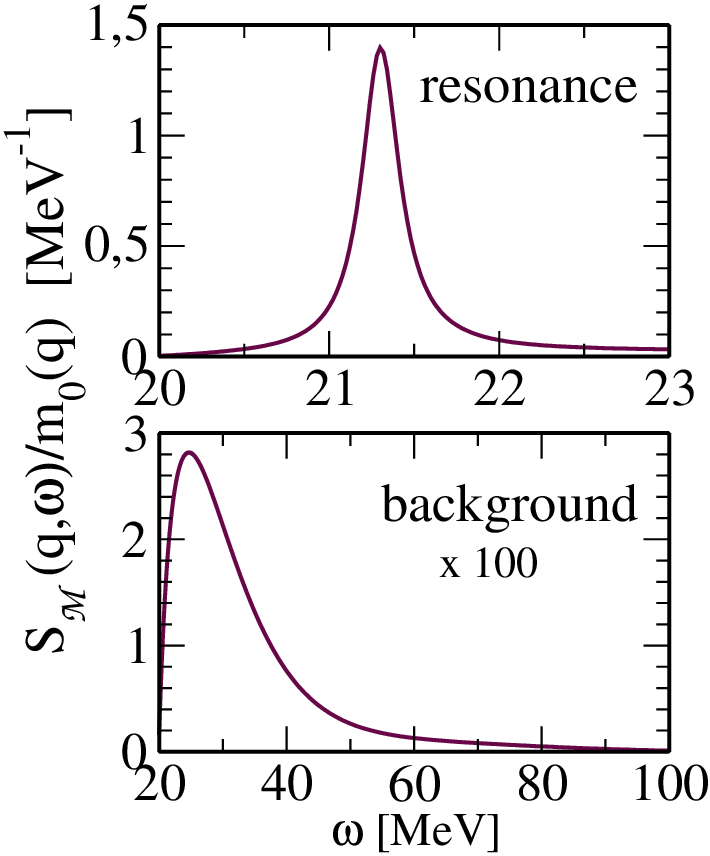}
\caption{Upper panel: resonance contributions to the inelastic isoscalar monopole strength of $^4$He 
as a function of the excitation energy; lower panel: background contribution magnified by a factor 100. }
\label{fig-2}       
\end{figure}

In~\cite{BaB13} the transition form factor to the $0^+$ excitation had been studied using the Lorentz Integral Transform 
(LIT) method~\cite{EfL94,EfL07} using the Effective Interaction Hyperspherical Harmonics (EIHH) expansion~\cite{EIHH}. In ~\cite{breath} 
we have focused on the isoscalar monopole strength distributions $S_M(q,\omega)$ for several fixed momentum transfers.
There it has been shown that especially at low momentum transfer (e.g. q=0.25 fm$^{-1}$) the isoscalar monopole spectrum 
is indeed dominated by a single resonant ''bump'' close to threshold, as it is evident in Fig.~\ref{fig-1}. 
Since the LIT method has allowed to separate the resonance contribution from the background (see Ref.~\cite{BaB13} for details) 
we show them separately here in Fig.~\ref{fig-2}. One can notice 
how far in energy the background extends, even if it remains very small. Figures 1 and 2 have been 
obtained with the chiral effective potential of Ref.~\cite{EnM03}, including both 2-body (at N3LO) and 3-body (at N2LO) contributions.

According to what was stated above, one can measure ''the degree of collectivity`` of the resonance state, 
by taking the ratio of the resonance area to the moment $m_0$. In this case one obtains 0.53. 
Values exceeding 50\% are generally considered an indication 
of a rather strong collectivity (see e.g.~\cite{Ro70}). Because 
of the extension of the background contribution at higher energy the ratio to $m_1$ is much smaller (34\%), and a conclusion about 
the concentration of strength in the resonance, would be 
misleading if compared to $S_M(q,\omega)$  as in Fig.~\ref{fig-1}, which shows a single prominent narrow peak in the spectrum.

\begin{figure}[htb]
\centering
\begin{center}
\includegraphics[width=7cm,clip]{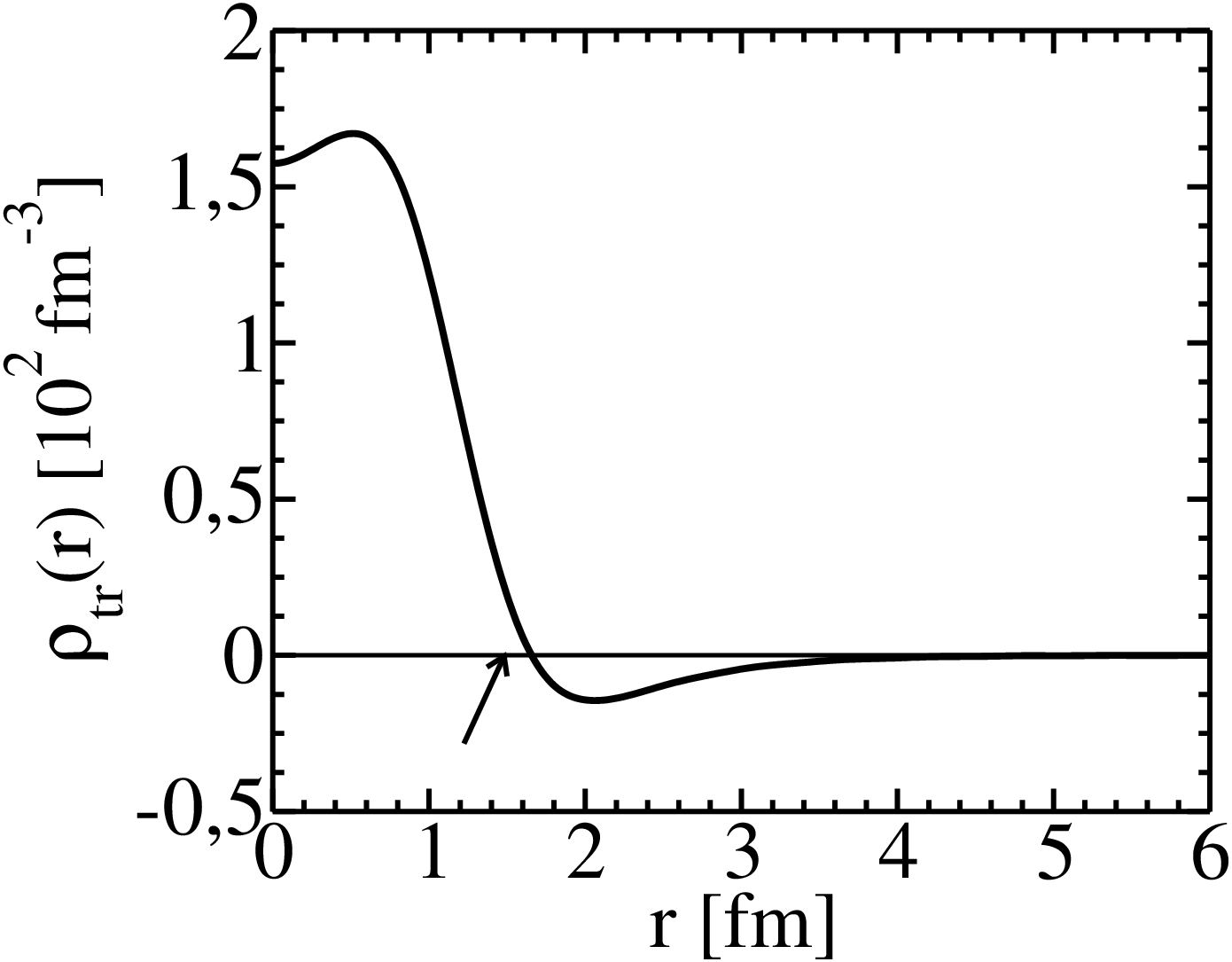}
\end{center}
\caption{The  transition form factor to the resonant state $0^+$. The arrow indicates the position of the root mean square radius.}
\label{fig-3}       
\end{figure}

In Fig.~\ref{fig-3} one can see the behavior of the transition density as a function of $r$. If one considers that 
the potential used in these calculations gives a root mean square radius of 1.46 fm one  notices 
that this does indeed correspond almost to the point where the transition cross section crosses the zero axis, 
a typical signature of some kind of breathing mode. 
The results shown in Figs. 1-3 seem to point out that the two criteria that are generally accepted as indications 
of a collective behavior are fulfilled in the case of the isoscalar monopole excitation of $^4$He. On the other hand it is clear that
if one observes this system with such a large wavelength one is likely to observe the dynamics of all nucleons together. In this case
it looks as if the resonance state $|R\rangle$ is obtained by a scaling transformation 
\begin{equation}
|R\rangle= e^{i \alpha \sum_i p_i r_i}|0\rangle=e^{i \lambda [T,M^{LW}]}|0\rangle
\end{equation}
where $M^{LW}=\sum_i r_i^2-\langle0| \sum_i r_i^2|0\rangle$ is proportional to the low-$q$ (long wavelength) limit of $M$. The transformation in the previous equation 
acts scaling the positions $r_i$ of all particles into $\alpha r_i$, and therefore also of the hyperradius, namely 
one of the six {\it collective} coordinates defined by the group $GL^{+}(3,R)$ 
~\cite{Zick71,Rowe80} (breathing).
In order to further prove that this is the case one could evaluate $m_{-1}$ (connected to the compressibility) in terms of 
the moment $m_3$, in a similar case as it was done in Ref.~\cite{LiO77}, and compare it to the value 
obtained integrating the inverse energy weighted strength of Fig.~\ref{fig-1}. Work in this direction is in progress.

\section*{References}

\end{document}